\documentclass[prd,showpacs,draft,amsfonts,amssymb,amsmath,nofootinbib]{revtex4}
\usepackage{graphicx}

\begin{document}
\title{Evolution Operators for Linearly Polarized Two-Killing Cosmological Models}

\author{J. Fernando \surname{Barbero G.}}
\email[]{fbarbero@iem.cfmac.csic.es} \affiliation{Instituto de
Estructura de la Materia, CSIC, Serrano 123, 28006 Madrid, Spain}
\author{Daniel  \surname{G\'omez Vergel}}
\email[]{dgvergel@iem.cfmac.csic.es} \affiliation{Instituto de
Estructura de la Materia, CSIC, Serrano 123, 28006 Madrid, Spain}
\author{Eduardo J. \surname{S. Villase\~nor}}
\email[]{ejsanche@math.uc3m.es} \affiliation{Grupo de
Modelizaci\'on y Simulaci\'on Num\'erica, Universidad Carlos III
de Madrid, Avda. de la Universidad 30, 28911 Legan\'es, Spain}
\affiliation{Instituto de Estructura de la Materia, CSIC, Serrano
123, 28006 Madrid, Spain}

\date{March 13, 2006}

\begin{abstract}
We give a general procedure to obtain non perturbative evolution
operators \textit{in closed form} for quantized linearly polarized
two Killing vector reductions of general relativity with a
cosmological interpretation. We study the representation of these
operators in Fock spaces and discuss in detail the conditions
leading to unitary evolutions.
\end{abstract}

\pacs{04.60.Ds, 04.60.Kz, 04.62.+v}

\maketitle

\section{Introduction}

The study of symmetry reductions has been a fruitful way to gain
valuable insights into the behavior of general relativity in its
quantum regime. This is so because they provide tractable models
where it is possible to make computations and obtain concrete
predictions, at least in a qualitative way, about relevant
features of a full blown theory of quantum gravity. Two Killing
vector reductions are specially appealing because they have local
degrees of freedom and (restricted) diffeomorphism invariance, two
of the features of general relativity that lie at the heart of the
difficulties encountered in its quantization. These reductions
differ from each other in the spacetime topology and the action of
the isometry group. When the Killing fields commute and are
hypersurface orthogonal the models become specially simple and it
is possible to solve them in closed form in a straightforward way.
In some cases the reduced model describes the propagation of
linearly polarized wave-like modes in a spacetime with non-compact
spatial slices, the so called Einstein-Rosen  waves
\cite{Einstein-Rosen}. Here the symmetry group is
$\mathbb{R}\times U(1)$ and the spacetime is topologically
$\mathbb{R}^4$.
In other situations, when the symmetry group is $U(1)\times U(1)$,
it provides cosmological models with initial singularities and an
assortment of spatial topologies: There are non-compact examples
such as the model introduced in \cite{Schmidt} (referred to in the
following as the Schmidt model) and compact ones such as the well
known Gowdy $T^3$ model \cite{Gowdy:1971jh,Gowdy:1973mu}, among
many others. All these symmetric sectors of vacuum general
relativity share the fact that the reduced phase space can be
parametrized by a scalar field in 1+1 dimensions and its
canonically conjugate momentum (plus, eventually, some
particle-like degrees of freedom). The main difference between
them, as far as the quantization is concerned, is that the
Hamiltonian for Einstein-Rosen waves \cite{Ashtekar:1996bb} is
time independent and has its origin in the surface terms needed to
have a well-defined action principle whereas the Hamiltonian in
the cosmological models is time dependent and is obtained through
deparametrization \cite{Pierri:2000ri,Beetle:1998iu}.

The quantization of linear Einstein-Rosen waves (free
\cite{Kuchar:1971xm,Ashtekar:1996bb,Ashtekar:1994ds,Ashtekar:1996yk}
or coupled to matter \cite{BarberoG.:2005ge}) is fairly well
understood. The quantum unitary evolution operator can be obtained
in closed form in a straightforward way and can be used for a
number of purposes leading to physical applications such as the
discussion of the existence of large quantum gravity effects
\cite{Ashtekar:1996yk} or the study of the microcausality of the
system \cite{BarberoG.:2003ye}. Models of the cosmological type,
specially the Gowdy $T^3$ one, have been much harder to crack even
though they received a lot of attention already in the seventies
by Misner \cite{Misner}, Berger \cite{Berger:1973}, and other
authors. Although the first attempts to quantize the system were
largely successful \cite{Berger:1975kn} and key technical insights
were introduced already at these stage some features of the
formalism were not completely satisfactory. In particular, the
reliance on a Hilbert space built as a tensor product of infinite,
one-particle Hilbert spaces is problematic since it is known that
such a tensor product is not separable and the representation of
the canonical commutation relations that it provides is reducible
\cite{Wald}. A renewed interest in the quantization of the
Einstein-Rosen, Gowdy, and Schmidt models arose in the nineties
\cite{Ashtekar:1996bb,Ashtekar:1996yk,Ashtekar:1994ds,Beetle:1998iu,Pierri:2000ri}
when, instead of the Dirac approach previously used, the
quantization of the system was done by gauge fixing. At the time
it was thought that they could provide suitable testing grounds
for loop quantum gravity and address some general issues related
to the quantization of general relativity. However the finding
that there were problems with the unitary implementation of
dynamics in the Gowdy $T^3$ model (and also in the Schmidt case)
\cite{Torre:2002xt,Corichi:2002vy} was somewhat of a shock and was
perceived as a potential drawback to its use as a toy model for
quantum gravity. The situation has recently improved after Corichi
et. al. \cite{Corichi:2005jb, Corichi:2006xi} have shown the
existence of unitary evolution for a field closely related to the
scalar that usually encodes the local gravitational degrees of
freedom.

The purpose of the present paper in this context is to obtain
evolution operators \textit{explicitly in closed form} written in
terms of the basic objects, i.e. the field and momentum operators.
This is done before choosing a specific Hilbert space
representation for them. So, even if they are formal at this
stage, they offer the possibility to explore different choices for
the quantization of the system and, in particular, non-Fock
representations (when available). Owing to this fact it is
important to notice that our approach differs from the usual ones
that make use of the Fock space --built from the solutions to the
field equations-- and a choice of complex structure to select the
one-particle Hilbert space. As an application of our formalism we
will discuss the unitary implementability of the time evolution of
the system as done in \cite{Corichi:2005jb} by changing the basic
fields used to encode the physical degrees of freedom of the
model. This provides an alternative point of view on this problem.

The outline of the paper is the following. After this introduction
we review in section \ref{U1} the Hamiltonian formalism for the
$U(1)\times U(1)$ symmetric models considered in the paper;
specifically we will discuss their derivation from an action
principle, gauge fixing, and deparametrization. After this we will
devote section \ref{evol} to obtain formal unitary evolution
operators $\hat{U}(t,t_0)$, defined in terms of abstract field and
momentum. This will be done in a unified way for a family of time
dependent Hamiltonians that include the Schmidt and Gowdy models.
Under the condition that the above mentioned formal operators can
be made unitary by choosing a suitable representation for the
field on a Hilbert space they can be shown to satisfy the
evolution equation
$i\hbar\partial_t\hat{U}(t,t_0)=\hat{H}(t)\hat{U}(t,t_0)$ where
$\hat{H}(t)$ is the time dependent Hamiltonian of the system. The
technical details of the construction --that relies upon known
results concerning the quantization of the time-dependent harmonic
oscillator-- will be left to appendix \ref{appendixA}.

As an application of our scheme we discuss in section \ref{fock}
the problem of finding suitable Fock space representations for the
field and its momentum that give rise to unitary evolution. At
this stage the use of the auxiliary field considered in
\cite{Berger:1973} and reintroduced in
\cite{Corichi:2005jb,Corichi:2006xi} will play an important role
that will be clarified within the present framework. We end the
paper with our conclusions in section \ref{concl} and three
appendices where we provide mathematical details on the obtention
of the evolution operators and related topics.

\section {Linearly Polarized $U(1)\times U(1)$ Models: Hamiltonian Formulation}{\label{U1}}

The Gowdy and Schmidt models have been extensively used to study
issues in quantum cosmology. In this context they are interesting
because they share with the full theory of general relativity some
of the features that renders its quantization highly non-trivial,
in particular, they are genuine field theories with (some)
diffeomorphism invariance. They have also been used to discuss
physical issues such as the quantum behavior of the initial
singularity in a non-homogeneous setting
\cite{Berger:1973,Husain:1987am}.

\subsection{The Midi-Superspace}

We will start by reviewing here some relevant facts about the
Gowdy and Schmidt models, in particular, the symmetry reduction
process and the deparametrization needed to define a convenient
time-dependent Hamiltonian. The spacetimes
$(\mathcal{M},{^{{\scriptscriptstyle(4)}}g_{ab}})$ of the Schmidt
and Gowdy $T^{3}$ models \cite{Gowdy:1973mu,Schmidt} are vacuum
solutions to the Einstein equations characterized by their
isometry group, in this case $U(1)\times U(1)$. Once the group
action is properly defined (smooth, effective, proper, and with
\emph{no fixed points}\footnote{This condition rules out the
$\mathbb{S}^1\times \mathbb{S}^2$, $\mathbb{S}^3$, and lens space
topologies for the Gowdy model.}) on the connected and oriented
spatial sections $\Sigma^3$ of $\mathcal{M}=\mathbb{R}\times
\Sigma^3$ the spatial topology is fixed to be of the form
$\Sigma^3=\mathbb{X}\times \mathbb{S}^{1}\times \mathbb{S}^{1}$,
where $\mathbb{X}=\mathbb{R}$ for Schmidt and
$\mathbb{X}=\mathbb{S}^1$ for Gowdy
\cite{Chrusciel:1990zx,Berger:1994sm}.

We will focus here on the linearly polarized case where the
isometry group is generated by a pair of mutually orthogonal,
commuting, spacelike, and globally defined
\textit{hypersurface-orthogonal} Killing vector fields
$(\xi^{a},\sigma^{a})$. The symmetry of the system implies that
the Lie derivatives of the metric vanish: $
L_{\xi}{^{{\scriptscriptstyle(4)}}g_{ab}}=L_{\sigma}{^{{\scriptscriptstyle(4)}}g_{ab}}=0$.

With these assumptions it is possible to exactly solve the vacuum
Einstein equation
\begin{equation}\label{vacua}
^{{\scriptscriptstyle(4)}}R_{ab}=0,
\end{equation}
where $^{{\scriptscriptstyle(4)}}R_{ab}$ is the Ricci tensor
associated with the Levi-Civita connection
$^{{\scriptscriptstyle(4)}}\nabla_{a}$ compatible with the
4-metric $^{{\scriptscriptstyle(4)}}g_{ab}$. Since the Killing
fields are hypersurface orthogonal, the space of orbits
$M=\mathbb{R}\times\mathbb{X}\times\mathbb{S}^1$ defined by one of
them, say $\xi^{a}$, can be identified as an embedded hypersurface
in $\mathcal{M}$ which is everywhere orthogonal to the (closed)
orbits of $\xi^{a}$. The induced 3-metric
${^{{\scriptscriptstyle(3)}}g_{ab}}$ on $M$ can be written in
terms of $\xi_a:={^{{\scriptscriptstyle(4)}}}g_{ab}\xi^b$ and
$\lambda:={^{{\scriptscriptstyle(4)}}}g_{ab}\, \xi^{a}\xi^{b}>0$
 as
\begin{equation}
{^{{\scriptscriptstyle(3)}}g_{ab}}={^{{\scriptscriptstyle(4)}}g_{ab}}-\lambda^{-1}\xi_{a}\xi_{b}\,.
\end{equation}
Let $^{{\scriptscriptstyle(3)}}R_{ab}$,
$^{{\scriptscriptstyle(3)}}\nabla_{a}$, and
$^{{\scriptscriptstyle(3)}}\square$ be, respectively, the Ricci
tensor, the metric connection, and the d'Alembert operator
associated to $^{{\scriptscriptstyle(3)}}g_{ab}$. The equation
(\ref{vacua}) is then equivalent to the system
\cite{Geroch:1970nt}
\begin{equation}\label{system1}
\left.\begin{array}{l}^{{\scriptscriptstyle(3)}}\square \lambda =
\frac{1}{2\lambda}\,{^{{\scriptscriptstyle(3)}}g^{ab}}
\left({^{{\scriptscriptstyle(3)}}\nabla_{a}}\lambda\right)\left({^{{\scriptscriptstyle(3)}}\nabla_{b}}
\lambda\right),\\
\\
{^{{\scriptscriptstyle(3)}}R}_{ab}=\frac{1}{2\lambda} \,
{^{{\scriptscriptstyle(3)}}\nabla}_{a}{^{{\scriptscriptstyle(3)}}\nabla}_{b}\lambda
-\frac{1}{4\lambda^{2}}\left({^{{\scriptscriptstyle(3)}}\nabla}_{a}\lambda\right)
\left({^{{\scriptscriptstyle(3)}}\nabla}_{b}\lambda\right).\\
\end{array}\right.
\end{equation}
The remaining Killing vector field $\sigma^{a}$ satisfies
$L_{\sigma}{^{{\scriptscriptstyle(3)}}}g_{ab}=0,\,\,\,L_{\sigma}\lambda=0$.

It is well known that in order to further simplify the equations
(\ref{system1}) we can use the conformal transformation
\begin{equation}\label{conform}
g_{ab}:=\lambda \,{^{{\scriptscriptstyle(3)}}}g_{ab},
\end{equation}
that is well defined since $\xi^{a}$ never vanishes. Let $R_{ab}$,
$\nabla_{a}$, and $\square$ be, respectively, the Ricci tensor,
the metric connection and the d'Alembert operator associated to
the new 3-metric $g_{ab}$. After the conformal rescaling
(\ref{conform}), the system (\ref{system1}) is equivalent to
\begin{equation}\label{EKG}
\left.\begin{array}{l}\square \phi =0\,,\\
\\
R_{ab}=\left(\nabla_{a}\phi\right)\left(\nabla_{b}\phi\right),\\
\end{array}\right.
\end{equation}
where we have defined the field $\sqrt{2}\phi:=\log \lambda$.
Therefore, the symmetry reduced models considered here can be
thought of as (2+1)-general relativity coupled to a massless
scalar field $\phi$, with the additional symmetry defined by the
Killing field $\sigma^a$. They can be derived from the 2+1
Einstein-Hilbert action for gravity coupled to a massless scalar
\begin{equation}\label{stdAct}
  S[g, \phi] = \frac{1}{16\pi G_{3}} \int_{M}  \,| g|^{1/2} \, \left( R -
  g^{ab} \,\nabla_a \phi \,\nabla_b \phi \right) + \frac{1}{8\pi G_{3}}
  \oint_{\partial M }  \,|h|^{1/2} \, K,
\end{equation}
that we will use as the starting point for the Hamiltonian
formalism. The last term must be introduced in the non-compact
case in order to have a well defined variational principle and
involves an integration over the  2-dimensional asymptotic
boundary of $M$. Here $R$ is the scalar curvature of $g_{ab}$, $h$
and $K$ are, respectively, the determinant of the induced metric
and the trace of the second fundamental form of the boundary
$\partial M$, and $G_3$ is the Newton constant per unit length in
the direction of the symmetry orbits.

In order to respect the symmetry in the Hamiltonian analysis of
(\ref{stdAct}) it is important to notice that, owing to the
hypersurface orthogonality condition, the metric can be decomposed
as
\begin{equation}
g_{ab}=h_{ab}+\tau^{-2}\sigma_{a}\sigma_{b},
\end{equation} where $\sigma_a:=g_{ab}\sigma^b$,
$\tau^{2}:=\sigma_a\sigma^a$ is the area density of the symmetry
group orbits, and $h_{ab}$ denotes the induced metric of signature
$(-+)$ on the 2-dimensional manifolds --topologically
$\mathbb{R}\times\mathbb{X}$-- that are everywhere orthogonal to
the closed orbits of $\sigma^{a}$. The symmetry present in this
case implies that the gradient of $\tau$ is always timelike. We
introduce now a foliation of $\mathbb{R}\times \mathbb{X}$ defined
by spacelike level hypersurfaces of a suitable scalar function
$t$. We also define a dynamical vector field $t^{a}$ such that
$t^{a}\nabla_{a}t=1$. By using the unit, future-pointing,
timelike, normal vector field to the foliation $n^a$ and the unit
vector field $\hat{x}^a$ compatible with the chosen orientation
and tangent to each slice (topologically $\mathbb{X}$) it is
possible to write $t^{a}=Nn^a+N^{x}\hat{x}^a$, where $N$ and
$N^{x}$ are the lapse and shift functions. The 2-metric $h_{ab}$
can be written then as $h_{ab}=-n_an_b+\hat{x}_a\hat{x}_b$, where
$n_a$ and $\hat{x}_a$ are respectively $g_{ab}n^{b}$ and
$g_{ab}\hat{x}^b$. We introduce the additional non-unit vector
field $x^{a}=\mathrm{e}^{\gamma/2}\hat{x}^{a}$, where $\gamma$ is
an extra field, and impose the vanishing of the Lie brackets of
the set of vector fields $(t^{a},x^{a},\sigma^{a})$:
\begin{equation}
\left.\begin{array}{l}
\partial_{\sigma}N=\partial_{\sigma}N^{x}=\partial_{\sigma}\gamma=0,\\
\left[\sigma,\hat{x}\right]^{a}=[\sigma,n]^{a}=0,\\
N\mathrm{e}^{\gamma/2}[\hat{x},n]^{a}=-n^{a}\partial_{x}N
-\hat{x}^{a}\left(\partial_{x}N^{x}-
\partial_{t}\mathrm{e}^{\gamma/2}\right).
\end{array}\right.\,
\end{equation}
By doing this it is possible to construct a global coordinate
system for $M$: $(t,x,\sigma)\in (0,\infty)\times \mathbb{X}\times
\mathbb{S}^1$. In these coordinates
\begin{equation}\label{metric}
g_{ab}=(-N^{2}+\left(N^{x}\right)^{2})\nabla_{a}t\nabla_{b}t+
2\mathrm{e}^{\gamma/2}N^{x}\nabla_{(a}t\nabla_{b)}x+
\mathrm{e}^{\gamma}\nabla_{a}x\nabla_{b}x+\tau^{2}\nabla_{a}\sigma\nabla_{b}\sigma,
\end{equation}
where the scalar fields $N$, $N^{x}$, and $\gamma$ depend only on
$t$ and $x$. Thus, the midi-superspace under consideration
consists of five real-valued smooth functions
$(N,N^{x},\tau,\gamma,\phi)$ that depend only on $(t,x)$ and
satisfy the Einstein-Klein-Gordon field equations (\ref{EKG}) with
the metric (\ref{metric}). In the Gowdy model, these functions are
periodic in $x$. However, in the Schmidt model $x\in \mathbb{R}$
and we need to impose asymptotic conditions for the fields in the
limits $x\rightarrow \pm\infty$. Here we will make use of the same
conditions introduced in \cite{Beetle:1998iu}
\begin{equation}\label{mspFall}
   \left.\begin{array}{l} \gamma \rightarrow \gamma_\pm(t) + O(x^{-1}), \quad
    N      \rightarrow N_\pm(t) + O(x^{-1}),      \\
    \tau   \rightarrow \tau_\pm(t) + O(x^{-1}),   \quad
    N^x    \rightarrow O(x^{-1}),                 \\
    \phi   \rightarrow O(x^{-1}).\end{array}\right.                \\
\end{equation}
Due to these fall-off conditions for the fields the boundary
integral in the action (\ref{stdAct}) vanishes in the non-compact
case (it is trivially zero in the compact one).

Introducing the metric (\ref{metric}) in the action (\ref{stdAct})
 it is straightforward to write
it in the canonical form
\begin{equation}
S[g, \phi] = \int_{t_1}^{t_2} \left(\int_{\mathbb{X}}
\left(p_{\gamma}\dot{\gamma}+p_{\tau}\dot{\tau}+p_{\phi}\dot{\phi}\right)\mathrm{d}x-
H[N,N^{x}]\right)\,\mathrm{d}t,
\end{equation}
where the dot denotes $\partial_t$, $p_{\gamma}$, $p_{\tau}$, and
$p_{\phi}$ are the canonically conjugate momenta [with the
fall-off conditions inherited from (\ref{mspFall})], and
$H[N,N^{x}]$ is the Hamiltonian that can be written as the sum of
the first class constraints $C[N]$ and $C_x[N^x]$
\begin{equation}
H[N,N^{x}]=C[N]+C_x[N^x]
=\int_{\mathbb{X}}NC\,\mathrm{d}x+\int_{\mathbb{X}}N^{x}C_{x}\,\mathrm{d}x\,.
\end{equation}
Here
\begin{eqnarray*}
&&C:=\frac{1}{8G_{3}}\mathrm{e}^{-\gamma/2}\left(2\tau''-\gamma'\tau'
-(8G_{3})^{2}p_{\tau}p_{\gamma}
+\frac{(8G_{3})^{2}}{4\tau}p_{\phi}^{2}+\tau\phi'^{2}\right),\\
&&C_{x}:=\mathrm{e}^{-\gamma/2}\left(p_{\tau}\tau'-2p_{\gamma}'+
p_{\gamma}\gamma'+\phi'p_{\phi}\right)\,,
\end{eqnarray*}
and we have denoted $\partial_x$ by a prime. In the following we
will choose units such that $8G_3=1$ so that the previous
constraints become
\begin{eqnarray*}
&&C[N]=\int_{\mathbb{X}}
N\mathrm{e}^{-\gamma/2}\left(2\tau''-\gamma'\tau'-p_{\tau}p_{\gamma}
+\frac{1}{4\tau}p_{\phi}^{2}+\tau\phi'^{2}\right)\mathrm{d}x,\\
&&C_{x}[N^x]=\int_{\mathbb{X}}N^x\mathrm{e}^{-\gamma/2}\left(p_{\tau}\tau'-2p_{\gamma}'+
p_{\gamma}\gamma'+\phi'p_{\phi}\right)\mathrm{d}x,
\end{eqnarray*}
where $N$ and $N^x$ are arbitrary functions. As the Hamiltonian is
zero --regardless of $\mathbb{X}$-- we have no dynamics so in
order to proceed further we will deparameterize the theory and
introduce a suitable phase space variable to play the role of
time.

\subsection{Deparameterization and Reduced Phase Space}

Deparameterization has been discussed in general in many places
(see, for example, \cite{Barvinsky:1993jf}); for the system under
consideration we closely follow \cite{Beetle:1998iu,Pierri:2000ri}
where the reader is directed for details. The fact that symmetry
forces the area density of the group orbits $\tau$ to have
time-like gradient and the analogy of the present models with the
Einstein-Rosen waves\footnote{For Einstein-Rosen waves the area
density has a space-like gradient and the gauge fixing that
simplifies the interpretation of the model leads to use it as a
radial coordinate.} suggests that we choose $\tau$ as our time
variable by imposing
\begin{equation}\label{gFix}
    \tau' = 0 \quad\mbox{and}\quad p_\gamma + 1 = 0.
\end{equation}
The first condition implies that $\tau$ is constant on the spatial
slices of the foliation whereas the second basically means that the
same is true for $\dot\tau$. Note that the last condition is
essentially equivalent to $p_\gamma' = 0$ because in the non-compact
case we have imposed a fall-off behavior for $p_\gamma$
\cite{Beetle:1998iu} forcing it to asymptotically approach $-1$.
These conditions imply that all the degrees of freedom but a single
one --that will change from one slice to another and will become the
time variable of our model-- are eliminated. In the compact case one
must use a gauge fixing of the type $p_\gamma + p = 0$ for a
constant $p$. This takes care of a global degree of freedom that
must be considered (see \cite{Pierri:2000ri}). As we are mostly
interested in the field-theoretic aspects of quantization for these
systems we will drop this here by using the condition $p_\gamma +
1=0$ in both cases. At this point one can check that the gauge
fixing conditions (\ref{gFix}) are admissible by computing their
Poisson brackets with the constraints \cite{Beetle:1998iu}. One can
also see that there is just one first class constraint, say
$C[N=e^{\gamma/2}]$, which cannot be solved after fixing the gauge.
This is found by solving for the lapse and shift that give a zero
Poisson bracket of the Hamiltonian with the gauge fixing conditions.
In fact the choice $N = e^{\gamma/2}$ will be such that
$\{\tau,C[N=e^{\gamma/2}]\}=1$ and, hence, we can identify $\tau$
with the time parameter $t$ of the system.

In these models there is a quadratic conserved momentum
\begin{equation}
P=\int_{\mathbb{X}}p_{\phi}\phi^{\prime}\,\mathrm{d}x\label{P}
\end{equation}
that is forced to be zero in the compact case. As a consequence of
this the reduced phase space for the Gowdy model is non-linear.

In order to complete the characterization of the reduced phase space
of the deparameterized theory we must solve the first class
constraints together with the gauge-fixing conditions. When this is
done we find that the action of the system can be completely
expressed in terms of the field $\phi$ and its canonically conjugate
momentum $p_{\phi}$ to give
\begin{equation}\label{redAct}
 s[\phi,p_\phi] = \int_{t_1}^{t_2}\!\!  \int_\mathbb{X}
  \bigg[
    p_\phi \dot\phi - \bigg(\frac{p_\phi^2}{4t} +
    t \phi'^2\bigg) \bigg]\,\mathrm{d}x\mathrm{d}t.
\end{equation}
The field equations are equivalent to the Hamilton equations
derived from the time-dependent Hamiltonian
\begin{equation}
H(t)=\int_\mathbb{X}
  \bigg(\frac{p_\phi^2}{4t} +
    t \phi'^2\bigg)\,\mathrm{d}x.
    \label{Hamiltonian}
\end{equation}
Once these equations are solved --and the constraint has been
taken into account for Gowdy-- the full four dimensional metric
can be built from their solutions by following in reverse the
reduction process for the metric and using the fact that the 2+1
dimensional metric (\ref{metric}), defined in $(0,\infty)\times
\mathbb{X}\times \mathbb{S}^1$, becomes
$$
g_{ab}=\mathrm{e}^{\gamma}(-\nabla_{a}t\nabla_{b}t+\nabla_{a}x\nabla_{b}x)+
t^{2}\nabla_{a}\sigma\nabla_{b}\sigma
$$
where $\gamma'=p_{\phi}\phi^{\prime}$. This metric displays a
curvature singularity at $t=0$. Finally it is important to remark
\cite{Pierri:2000ri,Beetle:1998iu} that the reduced action
(\ref{redAct}) corresponds, precisely, to that of a massless
scalar field (independent of the $\sigma$ coordinate) on a fixed
background spacetime with a metric given by
\begin{equation}\label{bgMet}
  \overcirc g_{ab}=-\nabla_a t \, \nabla_b t + \nabla_a x \, \nabla_b
  x+t^2 \nabla_a \sigma \, \nabla_b \sigma.
\end{equation}
This metric is defined on a 3-manifold $(0,\infty)\times
\mathbb{X}\times \mathbb{S}^1$ and shows, again,  the singular
behavior of the model at $t=0$.

\section{Evolution operators\label{evol}}

This section is devoted to study the quantization of quadratic,
time dependent, field Hamiltonians such as (\ref{Hamiltonian}).
The main difficulty to obtain evolution operators in closed form
is due to their explicit time dependence that precludes us from
writing them as exponentials of the Hamiltonian. A first point to
notice is that some of the functions of time that appear in
(\ref{Hamiltonian}) --say the one in the $p_{\phi}^2$ term-- can
be eliminated by a simple redefinition of the time variable. Let
us consider, then, the problem of quantizing a system with a time
dependent Hamiltonian of the general form
\begin{equation}
H(t)=\int_{\mathbb{X}}\bigg(\frac{1}{4}\pi^2(x)+\Lambda(t)\pi(x)\varphi(x)+
\omega^2(t)\varphi^{\prime 2}(x)\bigg)\mathrm{d}x.
\label{hamil_gen}
\end{equation}
Here $\varphi(x)$ and $\pi(x)$ are a field and its canonically
conjugate momentum [$\{\varphi(x),\pi(y)\}=\delta(x,y)$] defined
on $\mathbb{X}$, that may be either the real line $\mathbb{R}$
(Schmidt) or the circle $\mathbb{S}^1$ (Gowdy). The prime denotes
the $x$-derivative. The time functions $\Lambda(t)$ and
$\omega(t)$ will be determined by the particular model considered.
We allow for cross terms --absent in (\ref{Hamiltonian})--
involving fields and momenta as they will appear in the discussion
of the unitarity problem. We diagonalize the Hamiltonian
(\ref{hamil_gen}) by writing\footnote{We purposely use the same
letter to represent a field and its transform.}
\begin{eqnarray}
\varphi(x)=\frac{1}{\sqrt{2\pi}}\int_{\mathbb{\tilde{X}}}
\cos\Big(kx-\frac{\pi}{4}\Big)\varphi(k)\,\mathrm{d}\mu(k),&&\label{trans_cos}\\
\pi(x)=\sqrt{\frac{2}{\pi}}\int_{\mathbb{\tilde{X}}}
\cos\Big(kx-\frac{\pi}{4}\Big)\pi(k)\,\mathrm{d}\mu(k),&&\nonumber
\end{eqnarray}
where $\varphi(k)$ and $\pi(k)$ are also canonically conjugate
[$\{\varphi(k),\pi(q)\}=\delta(k,q)$ with a Dirac or Kronecker
delta]. Here $\mathbb{\tilde{X}}$ denotes the real line for
$\mathbb{X}=\mathbb{R}$ and the integers $\mathbb{Z}$ when
$\mathbb{X}=\mathbb{S}^1$. The measure $\mathrm{d}\mu(k)$ simply
refers to the fact that the previous integral either extends to
$\mathbb{R}$ or becomes a sum over the integers respectively. In
terms of them we have
\begin{eqnarray}
H(t)=\frac{1}{2}\int_{\mathbb{\tilde{X}}}\bigg(\pi^2(k)
+2\Lambda(t)\pi(k)\varphi(k)\!+\!k^2\omega^2(t)\varphi^2(k)\bigg)\,\mathrm{d}\mu(k).
\label{hamil_k}
\end{eqnarray}
This somewhat unusual diagonalization has the advantage of
decoupling the different modes \textit{right from the start} and
avoids the problems encountered in \cite{Pierri:2000ri} with the
non-diagonal form of the Hamiltonian in terms of creation and
annihilation operators. As we can see this is a sum of time
dependent uncoupled Hamiltonians that are closely related to the
harmonic oscillator with a time dependent frequency (even when the
$\Lambda(t)\pi(k)\varphi(k)$ term is present). In fact, denoting
$\partial_t$ by a dot, the Hamilton equations derived from
(\ref{hamil_k}) lead to
$$\ddot{\varphi}(k,t)+\Omega^2(k,t){\varphi}(k,t)=0$$
with
\begin{equation}
\Omega^2(k,t):=k^2\omega^2(t)-\Lambda^2(t)-\dot{\Lambda}(t)\,,\label{Omega}
\end{equation}
that we will suppose to be positive in the following.

To quantize the model we promote the field and momentum to
\textit{formal} algebraic objects $\hat{\varphi}(k)$ and
$\hat{\pi}(k)$ satisfying the canonical commutation relations
$[\hat{\varphi}(k),\hat{\pi}(q)]=i\hbar\delta(k,q)$ and symmetrize
the cross term in (\ref{hamil_k}). In the main body of the paper
we take $\hbar=1$. This way we obtain
\begin{equation}
\hat{H}(t)=\frac{1}{2}\int_{\tilde{\mathbb{X}}}
\bigg(\hat{\pi}^{2}(k)+\Lambda(t)
\left[\hat{\pi}(k)\hat{\varphi}(k)+\hat{\varphi}(k)\hat{\pi}(k)\right]
+k^{2}\omega^2(t)\hat{\varphi}^{2}(k)\bigg)\mathrm{d}\mu(k).\label{qHam}
\end{equation}
Explicit solutions for the evolution operator for problems similar
to this have been known for a number of years (see, for example,
\cite{JMP,FGuasti:2003}). Using them (see appendix \ref{appendixA})
we can write down the formal quantum evolution operator
$\hat{U}(t,t_0)$ as the product
$$\hat{U}(t,t_0)=\hat{\mathcal{D}}(t,t_0)\hat{\mathcal{S}}(t,t_0)\hat{\mathcal{R}}
(t,t_0)$$
with
\begin{widetext}
\begin{eqnarray}
&&\hat{\mathcal{D}}(t,t_0):=\exp\bigg(-\frac{i}{2}\int_{\tilde{\mathbb{X}}}
\Big[\frac{\dot{\rho}(k,t_0)}{\rho(k,t_0)}-\frac{\dot{\rho}(k,t)}{\rho(k,t)}+
\Lambda(t)-\Lambda(t_0)\Big]\hat{\varphi}^{\,2}(k)\,\mathrm{d}\mu(k)\bigg),\label{D}\\
&&\hat{\mathcal{S}}(t,t_0):=\exp\Bigg(\frac{i}{2}\int_{\tilde{\mathbb{X}}}
\bigg[\log\frac{\rho(k,t_0)}{\rho(k,t)}\bigg]\bigg\{\hat{\varphi}(k)\bigg[\hat{\pi}(k)-
\left(\frac{\dot{\rho}(k,t_0)}{\rho(k,t_0)}-\Lambda(t_0)\right)\hat{\varphi}(k)\bigg]\nonumber\\
&&\hspace{8.4cm}+\bigg[\hat{\pi}(k)-
\left(\frac{\dot{\rho}(k,t_0)}{\rho(k,t_0)}-\Lambda(t_0)\right)\hat{\varphi}(k)\bigg]\hat{\varphi}(k)
\bigg\}\,\mathrm{d}\mu(k)\Bigg), \label{S}\\
&&\hat{\mathcal{R}}(t,t_0):=\exp\Bigg(-\frac{i}{2}\int_{\tilde{\mathbb{X}}}
\bigg[\int_{t_0}^t\frac{d\tau}{\rho^2(k,\tau)}\bigg]
\bigg\{\frac{\hat{\varphi}^{\,2}(k)}{\rho^2(k,t_0)}+\rho^2(k,t_0)\bigg[\hat{\pi}(k)-
\left(\frac{\dot{\rho}(k,t_0)}{\rho(k,t_0)}-\Lambda(t_0)\right)\hat{\varphi}(k)\bigg]^2\bigg\}
\,\mathrm{d}\mu(k)\Bigg).\label{R}
\end{eqnarray}
\end{widetext}
Here $\rho(k,t)$ is any solution to the Ermakov-Pinney
equation\footnote{The Ermakov-Pinney equation is of central
importance in the classical and quantum treatment of the harmonic
oscillator with time-dependent frequency. It has been used, for
example, in some field theoretic cosmological problems (see for
example \cite{Bertoni:1997qb,Finelli:1999dk}) and minisuperspace
models \cite{Finelli:1997cr,Hawkins:2001zx}. In this paper we use it
in the context of the Fock quantization of time dependent field
theories.}
\begin{eqnarray}
\ddot{\rho}(k,t)+\Omega^2(k,t)\rho(k,t)=\rho^{-3}(k,t)\label{EPeq}
\end{eqnarray}
with $\Omega^2(k,t)$ defined in (\ref{Omega}). It can be shown
(see appendix \ref{appendixB}) that $\rho(k,t)$ never vanishes and
$\hat{U}(t,t_0)$ is \textit{independent} of the choice of this
solution. It must be said that the previous way to factorize
$\hat{U}(t,t_0)$ is by no means unique although this form is
specially adapted to the discussion of its unitarity in the next
section.

If we give a representation for the fields $\hat{\varphi}(k)$ and
$\hat{\pi}(k)$ satisfying
$[\hat{\varphi}(k),\hat{\pi}(q)]=i\delta(k,q)$ and such that the
exponents (multiplied by $i$) in (\ref{D},\ref{S},\ref{R}) are
self-adjoint, it is straightforward to check that
$\hat{U}(t_0,t_0)=\mathbb{I}$,
$\hat{U}(t,t_0)^{-1}=\hat{U}(t,t_0)^\dag$,
$\hat{U}(t_2,t_0)=\hat{U}(t_2,t_1)\hat{U}(t_1,t_0)$, and
$i\partial_t\hat{U}(t,t_0)=\hat{H}(t)\hat{U}(t,t_0)$ where
$\hat{H}(t)$ is the quantum Hamiltonian (\ref{qHam}), i.e.
$\hat{U}(t,t_0)$ is the quantum evolution operator of the system
in this representation. The evolution of the field and momentum
operators in the Heisenberg picture can then be computed in a
straightforward way and shown to satisfy the classical equations
of motion as expected (see appendix \ref{appendixA}).

\section{Fock representations and unitary implementability
of time evolution\label{fock}}

\subsection{General Considerations}

As a first application of the previously derived formula we will
discuss here the problem of finding Fock space representations for
the field and momenta leading to unitary evolution operators. This
is important because of the known obstruction to the
implementability of time evolution for this model in some of the
descriptions considered in the past. Our results in this regard
strongly support the satisfactory resolution of this issue that
appears in \cite{Corichi:2005jb}.

Let us suppose that we take a Fock space $\mathcal{F}$ and write
(overbars denote complex conjugation)
\begin{eqnarray}
\hat{\varphi}(k)&=&f(k)\hat{a}_k+\bar{f}(k)\hat{a}^\dag_k\,,\quad
\hat{\pi}(k)=g(k)\hat{a}_k+\bar{g}(k)\hat{a}^\dag_k \label{f&g}
\end{eqnarray}
in terms of the annihilation and creation operators $\hat{a}_k$
and $\hat{a}^\dag_k$ with
$[\hat{a}_k,\hat{a}^\dag_q]=\delta(k,q)$. The functions $f$ and
$g$ must satisfy
\begin{equation}
f(k)\bar{g}(k)-\bar{f}(k)g(k)=i, \label{commu}
\end{equation}
[so that $[\hat{\varphi}(k),\hat{\pi}(q)]=i\delta(k,q)$] but are
otherwise arbitrary at this stage. We want to find $f$ and $g$
such that $\hat{\mathcal{D}}(t,t_0)$, $\hat{\mathcal{S}}(t,t_0)$,
$\hat{\mathcal{R}} (t,t_0)$ --with normal ordered exponents to
prevent the appearance of infinite phases-- are unitary and
differentiable in $t$. An efficient procedure to discuss this
issue has been put forward by Torre in \cite{Torre:2002xt} by
using the theory of unitary implementation of canonical
transformations. Let us consider a general quantum operator of the
following type
\begin{equation}\label{genoperator}
\exp\left(i\int_{\tilde{
\mathbb{X}}}\left[\lambda_{1}(k):\hat{\varphi}^{2}(k):+\lambda_{2}(k)
:\hat{\pi}^{2}(k):
+\lambda_{3}(k):\hat{\varphi}(k)\hat{\pi}(k)+\hat{\pi}(k)\hat{\varphi}(k)
:\right]\mathrm{d}\mu(k)\right),
\end{equation}
where $\lambda_{1,2,3}$ are real functions of $k$. Notice that
$\hat{\mathcal{D}}(t,t_0)$, $\hat{\mathcal{S}}(t,t_0)$, and
$\hat{\mathcal{R}}(t,t_0)$ are particular cases of it with a
parametric dependence on $t$ and $t_0$. Introducing (\ref{f&g}) in
(\ref{genoperator}), the operator can be written as
\begin{equation}
\exp\left(i\int_{\tilde{\mathbb{X}}}[\chi_{1}(k)\hat{a}_{k}\hat{a}_{k}
+\bar{\chi}_{1}(k)\hat{a}^{\dag}_{k}\hat{a}^{\dag}_{k}+2\chi_{2}(k)\hat{a}^{\dag}_{k}
\hat{a}_{k}]\,\mathrm{d}\mu(k)\right),
\end{equation}
where
$\chi_{1}:=\lambda_{1}f^{2}+\lambda_{2}g^{2}+2\lambda_{3}fg$, and
$\chi_{2}:=\lambda_{1}|f|^{2}+\lambda_{2}|g|^{2}+\lambda_{3}(f\bar{g}+\bar{f}g)$
is a real function. We want to know now if the exponent (times
$i$) defines a self-adjoint operator. For fixed values of $t$ and
$t_0$ this is done by studying the auxiliary dynamics --in a
fictitious time parameter $s$-- defined by the exponent taken as a
classical Hamiltonian
\begin{equation}
\label{f}
F=\int_{\tilde{\mathbb{X}}}\left[\chi_1(k)a_{k}a_{k}
+\bar\chi_{1}(k)\bar{a}_{k}\bar{a}_{k}
+2\chi_{2}(k)\bar{a}_{k}a_{k}\right]\mathrm{d}\mu(k),
\end{equation}
and using the results on quantum implementability that appear in
\cite{Torre:2002xt}. The modes $a_k$ and $\bar{a}_k$ in (\ref{f})
are defined by the classical field and momentum
${\varphi}(k)=f(k){a}_k+\bar{f}(k)\bar{a}_k$,
${\pi}(k)=g(k){a}_k+\bar{g}(k)\bar{a}_k$ and satisfy
$\{a_{k},\bar{a}_{q}\}=-i\delta(k,q)$. In practice it is
convenient to consider the evolution equations
$$
\frac{\mathrm{d}a_k}{\mathrm{d}s}=\{a_{k},F\}
=-2i\left(\chi_2(k)a_{k}+\bar{\chi}_1(k)\bar{a}_{k}\right)\,, \quad
\frac{\mathrm{d}\bar{a}_k}{\mathrm{d}s}=\{\bar{a}_{k},F\}
=2i\left(\chi_{2}(k)\bar{a}_{k}+\chi_{1}(k)a_{k}\right)
$$
that are equivalent to the second order equations
\begin{equation}
\frac{\mathrm{d}^2a_k}{\mathrm{d}s^2}=4(|\chi_{1}(k)|^{2}-\chi_{2}^{2}(k))a_{k}
=4\left(\lambda_{3}^{2}(k)-\lambda_{1}(k)\lambda_{2}(k)\right)a_k\,.\label{s-evolution}
\end{equation}
These are linear equations so their solutions have a linear
dependence on the initial conditions $a_k(s_0)$ and
$\mathrm{d}a_k/\mathrm{d}s(s_0)=-2i\left(\chi_2(k)a_{k}(s_0)+\bar{\chi}_1(k)\bar{a}_k(s_0)\right)$
and, hence, on $a_k(s_0)$ and $\bar{a}_k(s_0)$. In order to
guarantee unitary implementability it suffices to show that the
integral over $\tilde{\mathbb{X}}$ of the modulus squared of the
coefficient of $\bar{a}_k(s_0)$ that appears in the solution of
(\ref{s-evolution}) is convergent. Finally, in order to verify the
strong continuity of the transformation in the auxiliary parameter
$s$ we have to check that the following limit
\begin{equation}
\label{limit} \lim_{s\rightarrow
s_0}\int_{\tilde{\mathbb{X}}}|a_{k}(s)-a_{k}(s_{0})|^{2}\,\mathrm{d}\mu(k)=0\,
\end{equation}
holds for the solution  $a_k(s)$ of (\ref{s-evolution}) with
square summable  initial data $a_k(s_0)$. We will obtain now
general conditions that guarantee that the previously defined
operators $\hat{\mathcal{D}}(t,t_0)$, $\hat{\mathcal{S}}(t,t_0)$,
and $\hat{\mathcal{R}}(t,t_0)$ (\ref{D})-(\ref{R}) are unitary:

\begin{itemize}

\item $\hat{\mathcal{D}}(t,t_0)$ is a quantum operator of the form
(\ref{genoperator}) with $\lambda_{2,3}=0$ and $\lambda_{1}\neq 0$
given in terms of the solution to the Ermakov-Pinney equation
(\ref{EPeq}) by
\begin{eqnarray*}
\lambda_1(k)=\frac{1}{2}\left(\frac{\dot\rho(k,t)}{\rho(k,t)}
-\Lambda(t)\right)-\frac{1}{2}\left(\frac{\dot\rho(k,t_0)}{\rho(k,t_0)}-\Lambda(t_0)\right)\,.
\end{eqnarray*}
In this case the solution to (\ref{s-evolution}) is given by
$$
a_k(s)=[1-2i(s-s_0)\chi_2(k)] a_k(s_0)-2i(s-s_0)\bar{\chi}_1(k)
\bar{a}_k(s_0)
$$
with $\chi_1=\lambda_1 f^2$ and $\chi_{2}=\lambda_{1}|f|^{2}$.
Then, the unitarity condition is
\begin{equation}
\int_{\tilde{\mathbb{X}}} \lambda^2_1(k)|f(k)|^4
\,\mathrm{d}\mu(k)<\infty, \label{unitD}
\end{equation}
and the strong continuity condition (\ref{limit}) becomes
\begin{eqnarray}
 0&=& \lim_{s\rightarrow
s_{0}}(s-s_{0})^2\int_{\tilde{\mathbb{X}}}\lambda_{1}^2(k)|\bar{f}^2(k)\bar{a}_k(s_0)+|f(k)|^2
a_k(s_0)|^2\,\mathrm{d}\mu(k). \label{unitD_bis}
\end{eqnarray}
This is
trivially satisfied whenever the last integral is well defined.\\

\item $\hat{\mathcal{S}}(t,t_0)$ is a quantum operator of the form
(\ref{genoperator}) with $\lambda_{2}= 0$, $\lambda_{1,3}\neq 0$:
\begin{eqnarray*}
\lambda_1(k)&=&-\left(\frac{\dot\rho(k,t_0)}{\rho(k,t_0)}-\Lambda(t_0)\right)\log\frac{\rho(k,t_0)}{\rho(k,t)},\\
\lambda_3(k)&=&\frac{1}{2}\log\frac{\rho(k,t_0)}{\rho(k,t)}.
\end{eqnarray*}
The solution to (\ref{s-evolution}) is now
\begin{eqnarray*}
a_k(s)&=&\{\cosh[2\lambda_3(k)(s-s_0)]
-i\chi_2(k)\lambda_3^{-1}(k)\sinh[2\lambda_3(k)(s-s_0)]\}\,
a_k(s_0)\\
&-&i\bar{\chi}_1(k)\lambda_3^{-1}(k)\sinh[2\lambda_3(k)(s-s_0)]\,
\bar{a}_k(s_0)\,,
\end{eqnarray*}
where $\chi_{1}=\lambda_{1}f^{2}+2\lambda_{3}fg$, and
$\chi_{2}=\lambda_{1}|f|^{2}+\lambda_{3}(f\bar{g}+\bar{f}g)$.

The unitarity condition in this case is
\begin{equation}
\int_{\tilde{\mathbb{X}}} |f(k)|^2
|2g(k)+\lambda_1(k)f(k)\lambda_3^{-1}(k)|^2
\sinh^2[2\lambda_3(k)(s-s_{0})]\,\mathrm{d}\mu(k)<\infty.
\label{unitS}
\end{equation}
It is important to point out here that the condition (\ref{commu})
implies that $|f(k)|^2|g(k)|^2\geq1/4$ and, hence, the convergence
of (\ref{unitS}) requires an appropriate fall-off of $|1+\lambda_1
f/(2\lambda_3g)|^2\sinh^2[2\lambda_3 (s-s_{0})]$. The strong
continuity condition is now
\begin{eqnarray}
\nonumber\hspace{-1cm}0&\!\!\!=\!\!\!\!&\lim_{s\rightarrow
s_{0}}\!\int_{\tilde{\mathbb{X}}}|\{\cosh
[2\lambda_{3}(k)(s-s_{0})]-i[
f(k)\bar{g}(k)+\bar{f}(k)g(k)+\lambda_{1}(k)\lambda^{-1}_{3}(k)|f(k)|^{2}
]\sinh [2\lambda_{3}(k)(s-s_{0})]-1\}  a_{k}(s_{0})
\\
&&\hspace{1.5cm}-i[\lambda_{1}(k)\lambda^{-1}_{3}(k)\bar{f}^{2}(k)+2\bar{f}(k)\bar{g}(k)]
\sinh[2\lambda_{3}(k)(s-s_{0})] \bar{a}_{k}(s_{0})
|^{2}\,\mathrm{d}\mu(k).\label{unitS_bis}
\end{eqnarray}

\item Finally $\hat{\mathcal{R}}(t,t_0)$ is a quantum operator of
the form (\ref{genoperator}) with $\lambda_{1,2,3}\neq 0$ given by
\begin{eqnarray*}
\lambda_1(k)&=&-\frac{1}{2}\left[\frac{1}{\rho^2(k,t_0)}+\rho^2(k,t_0)
\left(\frac{\dot{\rho}(k,t_0)}{\rho(k,t_0)}-\Lambda(t_0)\right)^2\right]\int_{t_0}^t
\frac{d\tau}{\rho^2(k,\tau)}\,,\\
\lambda_2(k)&=&-\frac{\rho^2(k,t_0)}{2}\int_{t_0}^t
\frac{d\tau}{\rho^2(k,\tau)}\,,\\
\lambda_3(k)&=&\frac{\rho^2(k,t_0)}{2}\left(\frac{\dot\rho(k,t_0)}{\rho(k,t_0)}-\Lambda(t_0)\right)\int_{t_0}^t
\frac{d\tau}{\rho^2(k,\tau)}\,,\\
\mathrm{with} \quad
\lambda^2(k)&:=&\lambda_1(k)\lambda_2(k)-\lambda_3^2(k)=\frac{1}{4}\left(\int_{t_0}^t
\frac{d\tau}{\rho^2(k,\tau)}\right)^2>0\,.\\
\end{eqnarray*}
The solution to (\ref{s-evolution}) is given by
\begin{eqnarray*}
a_k(s)&=&\{\cos[2\lambda(k)(s-s_0)]
-i\chi_2(k)\lambda^{-1}(k)\sin[2\lambda(k)(s-s_0)]\}\,
a_k(s_0)\\
&-&i\bar{\chi}_1(k)\lambda^{-1}(k)\sin[2\lambda(k)(s-s_0)]\,
\bar{a}_k(s_0)\,,
\end{eqnarray*}
with $\chi_{1}=\lambda_{1}f^{2}+\lambda_{2}g^{2}+2\lambda_{3}fg$,
and
$\chi_{2}=\lambda_{1}|f|^{2}+\lambda_{2}|g|^{2}+\lambda_{3}(f\bar{g}+\bar{f}g)$.

Then, the unitarity condition is
\begin{equation}
\int_{\tilde{\mathbb{X}}}
|\lambda_{1}(k)f^{2}(k)+\lambda_{2}(k)g^{2}(k)
+2\lambda_{3}(k)f(k)g(k)|^2\lambda^{-2}(k)\sin^2[2\lambda(k)
(s-s_{0})]\,\mathrm{d}\mu(k)<\infty, \label{unitR}
\end{equation}
and the strong continuity condition is
\begin{eqnarray}
\nonumber \hspace{-1cm} 0&=& \lim_{s\rightarrow
s_{0}}\int_{\tilde{\mathbb{X}}}|
\{[\lambda_1(k)|f(k)|^2+\lambda_2(k)|g(k)|^2
+\lambda_3(k)(f(k)\bar{g}(k)+\bar{f}(k)g(k))]\sin
[2\lambda(k)(s-s_{0})] \\
\nonumber &&\hspace{5.9cm}+i\lambda(k)(\cos
[2\lambda(k)(s-s_{0})]-1)\}a_{k}(s_{0})
\\
&&\hspace{1cm}+
[\lambda_1(k)\bar{f}^{2}(k)+\lambda_{2}(k)\bar{g}^{2}(k)
+2\lambda_3(k)\bar{f}(k)\bar{g}(k)]\sin[2\lambda(k)(s-s_{0})]
\bar{a}_{k}(s_{0})|^{2}\,\lambda^{-2}(k)\mathrm{d}\mu(k)\,.\label{unitR_bis}
\end{eqnarray}

\end{itemize}

\subsection{The Gowdy and Schmidt Models\label{GS}}

As we have previously shown in section \ref{U1}, the reduced
Hamiltonian (\ref{Hamiltonian}) for the Gowdy and Schmidt models
is
$$
H(t)=\int_{\mathbb{X}}\bigg(\frac{p_{\phi}^2(x)}{4t}+t
\phi^{\prime 2}(x)\bigg)\mathrm{d}x.
$$
After redefining the time parameter as $t=e^T$ and using the
cosine transform (\ref{trans_cos}) we get
\begin{equation}
H(T)=\frac{1}{2}\int_{\tilde{\mathbb{X}}}\bigg(p_{\phi}^2(k)+k^2e^{2T}
\phi^2(k)\bigg)\,\mathrm{d}\mu(k). \label{Ham_Sch_phi_k}
\end{equation}
In both cases there is a conserved quantity (\ref{P}) of the form
$$
P=\int_{\tilde{\mathbb{X}}}kp_{\phi}(-k)\phi(k)\,\mathrm{d}\mu(k).
$$
For the Gowdy model $P$ is constrained to be zero. This is usually
taken into account in the quantum theory as a condition on the
physical states that is trivially preserved under the time evolution
defined by the Hamiltonian (whenever it is well defined).

We discuss now the unitarity of the time evolution defined by the
explicit operator $\hat{U}(T,T_0)$ introduced above. The
Hamiltonian (\ref{Ham_Sch_phi_k}) belongs to the general class
(\ref{hamil_k}) with $\Lambda(T)=0$ and $\omega^2(T)=e^{2T}$. Then
(\ref{Omega}) gives $\Omega^2(k,T)=k^2e^{2T}$ and, in this case,
the general solution to the Ermakov-Pinney equation (\ref{EPeq})
can be written in terms of Bessel functions as
$$\rho(k,T)=[A J_0^2(|k|e^T)+2B
J_0(|k|e^T)Y_0(|k|e^T)+CY_0^2(|k|e^T)]^{1/2}
$$
with $AC-B^2=\pi^2/4$. It is possible to understand the difficulties
to get a unitary evolution operator as the impossibility of
fulfilling the unitarity conditions for some of the operators used
to build $\hat{U}(T,T_0)$, in particular the condition (\ref{unitS})
for $\hat{\mathcal{S}}(T,T_0)$. To show this we look at the leading
asymptotic behavior of
$\lambda_3(k)=(1/2)\log[\rho(k,T_0)/\rho(k,T)]$ when
$|k|\rightarrow\infty$ that is given by $(T-T_0)/4$ plus a bounded
periodic function of $|k|$. We have also
$\lambda_1(k)/\lambda_3(k)=-2[\dot{\rho}(k,T_{0})/\rho(k,T_{0})-\Lambda(T_{0})]\sim
1$. This shows that the convergence of  (\ref{unitS}) would require
$\lim_{|k|\rightarrow\infty} f(k)/g(k)=-2$ which is not compatible
with (\ref{commu}) and, hence, $\hat{\mathcal{S}}(T,T_0)$ cannot be
unitary.

The way this factor fails to be unitary suggests that one could
avoid this problem by introducing from the start (say at the
Lagrangian level and, hence, before obtaining the Hamiltonian) a
new field $\xi$ differing from $\phi$ in a certain time dependent
factor. The reason is to modify the leading asymptotics of
$\lambda_3(k)$ to improve the convergence of (\ref{unitS}). A
possible way to change that behavior within the conceptual scheme
that we are using here is to factorize the scalar field
$\phi(k,T)=h(T)\xi(k,T)$ for an appropriately chosen function
$h(T)$. It is always possible to find a unique (modulo a
multiplicative constant) time redefinition for a given $h$ in such
a way that the new field $\xi$ satisfies, again, a differential
equation for a harmonic oscillator with time (and $k$) dependent
frequency. It could be possible, in principle, to obtain general
conditions for $h$ that guarantee that the function $\lambda_3(k)$
has the right asymptotic behavior in $k$. Physically one would
expect that a choice for which the frequency squared is a sum of
$k^2$ plus a decreasing function of time would work as the system
would approach a free one for which the evolution operator
$\hat{U}$ in the form written above is well defined and unitary.
The essentially unique way to do this (see appendix
\ref{appendixC}) is to introduce a new field $\xi$ satisfying
\begin{equation}
\xi(x,t)=\sqrt{t}\phi(x,t) \label{xi}
\end{equation}
as was done in \cite{Berger:1973,Corichi:2005jb}. Here the
appropriate time variable is precisely the original one $t=e^T$.
The Hamiltonian in terms of $\xi(x)$ and its canonically conjugate
momentum $p_{\xi}(x)$ becomes \cite{Berger:1973}
\begin{equation}
H(t)=\int_{\mathbb{X}}\bigg(\frac{p_{\xi}^2(x)}{4}+\frac{\xi(x)p_{\xi}(x)}{2t}+
\xi^{\prime 2}(x)\bigg)\mathrm{d}x. \label{Ham_xi}
\end{equation}
Notice that this is \emph{not} the Hamiltonian considered in
\cite{Corichi:2005jb}. Although it is related to it by a canonical
transformation, the two \textit{quantum} dynamics are different
because the same wave function at an initial time $t_0$ evolves
differently. Diagonalizing (\ref{Ham_xi}) by using the transform
(\ref{trans_cos}), defined now for $\xi$ and $p_{\xi}$, and
introducing the corresponding operators $\hat{\xi}(k)$ and
$\hat{p}_{\xi}(k)$ we get
$$
\hat{H}(t)=\frac{1}{2}\int_{\tilde{\mathbb{X}}}\bigg(\hat{p}_{\xi}^2(k)
+\frac{1}{2t}[\hat{p}_{\xi}(k)\hat{\xi}(k)+\hat{\xi}(k)\hat{p}_{\xi}(k)]+
k^2\hat{\xi}^2(k)\bigg)\,\mathrm{d}\mu(k).
$$
As we can see the Hamiltonian  (\ref{Ham_xi}) belongs to the class
(\ref{hamil_gen}) with $\Lambda(t)=1/(2t)$, $\omega^2(t)=1$. In this
case, by using (\ref{Omega}), $\Omega^2(k,t)=k^2+1/(4t^2)$ and a
solution to the corresponding Ermakov-Pinney equation (\ref{EPeq})
is
\begin{equation}
\rho(k,t)=\sqrt{\pi t/2}[J_0^2(|k|t)+Y_0^2(|k|t)]^{1/2}.
\label{ro}
\end{equation}
The asymptotic behavior of the functions associated to this
solution that appear in the unitarity conditions
(\ref{unitD},\ref{unitS},\ref{unitR}) is given in the following
table
\begin{center}
\begin{tabular}{|c|l|l|}
\hline\hline
\vspace*{-4mm}& & \\
Function & $\hspace{.5cm}\displaystyle|k|\rightarrow\infty$ & $\hspace{1.3cm}\displaystyle|k|\rightarrow0$\\
\vspace*{-4mm}& & \\
\hline\hline
\vspace*{-4mm}& & \\
$\left.\vspace*{1mm}\displaystyle\rho(k,t)\right.$&
$\displaystyle\sim\frac{1}{\sqrt{|k|}}$ &$\displaystyle\sim
\sqrt{\frac{2t}{\pi}}\log(|k|t)$\\
\vspace*{-4mm}& & \\
\hline
\vspace*{-4mm}& & \\
$\displaystyle\int_{t_0}^t\frac{d\tau}{\rho^2(k,\tau)}$&
$\displaystyle\sim C(t,t_0)|k|$ &
$\displaystyle\sim\frac{\log(t/t_0)}{\log^2|k|}$ \\\vspace*{-4mm}& & \\
\hline \vspace*{-4mm}&
&\\
$\displaystyle\log{\frac{\rho(k,t_0)}{\rho(k,t)}}$&$\displaystyle\sim
\frac{t_0^2-t^2}{16|k|^2t^2t_0^2}$ &$\displaystyle\sim\log(t_0/t)
\left[\frac{1}{2}+\frac{1}{\log|k|}\right]$\\\vspace*{-4mm}& & \\
\hline\vspace*{-4mm}& &
\\$\displaystyle\frac{\dot{\rho}(k,t)}{\rho(k,t)}-\Lambda(t)$&$\displaystyle\sim
\frac{1}{8|k|^2t^3}-\frac{1}{2t}$
&$\displaystyle\sim\frac{1}{t\log(|k|t)}_{\,_{\,_{\,_{\,}}}}$\\
\hline\hline
\end{tabular}
\end{center}
where we do not need the specific form of $C(t,t_0)$. They can be
easily derived from (\ref{ro}) with the exception of the second one
that requires the use of equation (\ref{fin}) of appendix
\ref{appendixB}. Choosing, for example,
$$f(k)=1/\sqrt{2(1+|k|)}\quad \mathrm{and}\quad
g(k)=-i\sqrt{(1+|k|)/2}$$ it is straightforward to check that the
unitarity conditions (\ref{unitD},\ref{unitS},\ref{unitR}) are
satisfied together with (\ref{commu}). Finally, strong continuity
in the auxiliary parameter $s$ follows from
(\ref{unitD_bis},\ref{unitS_bis},\ref{unitR_bis}), by using the
asymptotic expansions given above and the fact that $a_k(s_0)$ is
square summable. We conclude, hence, that the exponents (times
$i$) in $\hat{\mathcal{D}}(t,t_0)$, $\hat{\mathcal{S}}(t,t_0)$,
and $\hat{\mathcal{R}}(t,t_0)$ are self-adjoint and these are
unitary.

In order to verify that $\hat{U}(t,t_0)$ satisfies the evolution
equation one must be able to compute derivatives in $t$. This can
be shown to be formally possible due to the fact that $t$
derivatives of the exponent trivially commute with it. Also, one
can see (appendix \ref{appendixA}) that the evolution of the field
and momentum operators generated by $\hat{U}(t,t_0)$ is
differentiable and their time derivatives are obtained as
commutators with the Hamiltonian. This allows us to make sense of
the time derivative of $\hat{U}(t,t_0)$ indirectly through its
action on the basic operators. Of course, one can try to study the
behavior in $(t,t_0)$ in a rigorous mathematical sense from first
principles but this is beyond the scope of the present paper.

\section{Conclusions and Comments\label{concl}}

Some comments are in order now. First of all it must be said that if
one is only interested in  Fock space representations the evolution
of $n$-particle states can be simply derived from the evolution of
creation operators in the Heisenberg picture (that can be directly
read from the classical dynamics of the system) and the evolution of
the vacuum state. Furthermore, the vacuum evolution can be written
in closed form as in \cite{Torre:1997zs,Pilch:1987eb}. This provides
the matrix elements of the evolution operator in a concrete Hilbert
space basis. A potential advantage of the framework discussed in
this paper is that one can, in principle, try to use non-Fock
representations to define the evolution operator given in closed
form by (\ref{D},\ref{S},\ref{R}). A second comment is that
$$\frac{\hat{\xi}(x)}{\sqrt{t}}=\frac{1}{\sqrt{2\pi
t}}\int_{\tilde{\mathbb{X}}}
[f(k)\hat{a}_k+\bar{f}(k)\hat{a}^\dag_k]\cos\Big(kx-\frac{\pi}{4}\Big)\,
\mathrm{d}\mu(k),$$ interpreted as an \textit{explicitly} time
dependent operator in the Schr\"{odinger picture} evolves, in the
Heisenberg picture, as the classical scalar field $\phi(x,t)$ that
encodes the physical degrees of freedom of the Gowdy and Schmidt
models. This evolution is perfectly well defined and unitary. As
we can see the problem is not that it is impossible to define a
unitary evolution for an object that behaves as the scalar field
$\phi$ but rather the \emph{impossibility} of doing this with a
representation of the type
$$
\hat{\phi}(k)=f(k)\hat{a}_k+\bar{f}(k)\hat{a}^\dag_k\,,\quad
\hat{p}_\phi(k)=g(k)\hat{a}_k+\bar{g}(k)\hat{a}^\dag_k\, .
$$
We want to emphasize that the use of the field $\xi$ defined in
(\ref{xi}) is as justified as that of $\phi$ because it can be
chosen as the fundamental object already at the Lagrangian level.
It is important, however, to realize that the quantum dynamics
must then be obtained from the corresponding Hamiltonian. In this
respect it should be noticed that it is possible to write down the
Hamiltonian corresponding to the dynamics considered in
\cite{Corichi:2005jb}, build the unitary evolution operator in
that case with our methods.

The possible uses of the approach that we have presented here are
manifold. In addition to the discussion of consistency issues
related to the unitarity of the time evolution of the system other
problems that may be tackled are the generalization of the
coherent and squeezed states for the harmonic oscillator with time
dependent frequency to these quantum cosmological models, the
study of particle creation, and the introduction  and discussion
of new representations to quantize the system.

\begin{acknowledgments}
The authors want to thank J. Cortez, I. Garay, L. Garay, J. M.
Mart\'{\i}n Garc\'{\i}a, G. Mena Marug\'an, and M. Varadarajan for
their insightful comments. This work was supported in part by the
Spanish MEC grant BFM2002-04031-C02-02 and D. G\'omez Vergel
acknowledges the financial support provided by CSIC through the
Introduction to Research assistantship and I3P programs.
\end{acknowledgments}

\begin{appendix}

\section{Computing the Unitary Evolution
Operator\label{appendixA}}

We give here some details about the construction of the formal
evolution operators for the kind of system considered in the
paper. In order to quantize the model defined by the Hamiltonian
(\ref{hamil_k}), we promote the field and momentum to formal
operators $\hat{\varphi}(k)$, $\hat{\pi}(k)$, satisfying the
canonical commutation relations
$[\hat{\varphi}(k),\hat{\pi}(q)]=i\hbar\delta(k,q)$, and
symmetrize the cross term in the classical Hamiltonian to arrive
at (\ref{qHam}). We want to solve now the Schr\"{o}dinger
equation\footnote{We assume that a suitable Hilbert space
representation exists.} $i\hbar
\partial_{t}|\psi(t)\rangle=\hat{H}(t)|\psi(t)\rangle$.
The strategy that we follow is to generalize the results already
known for a single harmonic oscillator with a time dependent
frequency to an infinite system of uncoupled harmonic oscillators
\cite{JMP,FGuasti:2003}. In order to obtain the quantum evolution
operator, we define
\begin{eqnarray}
&&\hspace{-6mm}\hat{D}(t;f):=\mathrm{exp}\left(-\frac{i}{2\hbar}\int_{\tilde{\mathbb{X}}}
f(k,t)\hat{\varphi}^2(k)\,\mathrm{d}\mu(k)\right),\nonumber\\
&&\hspace{-6mm}\hat{S}(t;g):=\mathrm{exp}\left(\frac{i}{2\hbar}\int_{\tilde{\mathbb{X}}}
g(k,t)\!\left[\hat{\varphi}(k)\hat{\pi}(k)+\hat{\pi}(k)\hat{\varphi}(k)\right]\,\mathrm{d}\mu(k)\right)\nonumber
\end{eqnarray}
depending on some functions $f$ and $g$ that will be fixed in the
following. Let us consider their product
$\hat{T}(t;f,g)=\hat{S}(t;g)\hat{D}(t;f)$ and introduce the state
vector $|\chi(t;f,g)\rangle=\hat{T}(t;f,g)|\psi(t)\rangle$ written
in terms of a solution to the Schr\"{o}dinger equation
$|\psi(t)\rangle$. We find now the equation that
$|\chi(t;f,g)\rangle$ must satisfy and choose the functions
$f(k,t)$ and $g(k,t)$ in such a way that the evolution for this
vector can be easily obtained
\begin{widetext}
\begin{eqnarray}
i\hbar\partial_{t}|\chi(t;f,g)\rangle&=&\bigg(\hat{T}(t;f,g)\hat{H}(t)\hat{T}^{\dag}(t;f,g)
-i\hbar\hat{T}(t;f,g)\partial_{t}\hat{T}^{\dag}(t;f,g)\bigg)|\chi(t;f,g)\rangle
\nonumber\\
&=&\frac{1}{2}\int_{\tilde{\mathbb{X}}}\mathrm{d}\mu(k)\bigg(\mathrm{e}^{-2g(k,t)}\hat{\pi}^2(k)
+\Big[f(k,t)+\Lambda(t)-\dot{g}(k,t)\Big]
\Big[\hat{\varphi}(k)\hat{\pi}(k)+\hat{\pi}(k)\hat{\varphi}(k)\Big]\label{longeq}\\&&+
\mathrm{e}^{2g(k,t)}\Big[f^2(k,t)+k^2\omega^2(t)+2\Lambda(t)f(k,t)+\dot{f}(k,t)\Big]
\hat{\varphi}^2(k)\bigg) \, |\chi(t;f,g)\rangle\nonumber
\end{eqnarray}
\end{widetext}
where the dot denotes $\partial_{t}$ and we have used the
following relations satisfied by $\hat{D}(t;f)$ and $\hat{S}(t;g)$
\begin{eqnarray*}
&&\hat{D}(t;f)\hat{\varphi}(k)\hat{D}^{\dag}(t;f)=\hat{\varphi}(k),\\
&&\hat{D}(t;f)\hat{\pi}(k)\hat{D}^{\dag}(t;f)=\hat{\pi}(k)+f(k,t)\hat{\varphi}(k),\\
&&\hat{S}(t;g)\hat{\varphi}(k)\hat{S}^{\dag}(t;g)=\hat{\varphi}(k)\mathrm{e}^{g(k,t)},\hspace{1.2cm}\\
&&\hat{S}(t;g)\hat{\pi}(k)\hat{S}^{\dag}(t;g)=\hat{\pi}(k)\mathrm{e}^{-g(k,t)}.\hspace{1.2cm}
\end{eqnarray*}
The form of (\ref{longeq}) suggests to choose functions $f$, $g$
such that
\begin{eqnarray*}
&&\hspace{-1cm} f(k,t)+\Lambda(t)=\dot{g}(k,t),\\
&&\hspace{-1cm}
f^2(k,t)+k^2\omega^2(t)+2\Lambda(t)f(k,t)+\dot{f}(k,t)=\mathrm{e}^{-4g(k,t)}.
\end{eqnarray*}
The equation for $|\chi(t)\rangle$ simplifies then to
\begin{eqnarray*}
i\hbar\partial_{t}|\chi(t)\rangle=
\frac{1}{2}\int_{\tilde{\mathbb{X}}}\mathrm{d}\mu(k)\,\,\mathrm{e}^{-2g(k,t)}
\big[\hat{\pi}^2(k)+\hat{\varphi}^2(k)\big] |\chi(t)\rangle.
\end{eqnarray*}
A concrete way to achieve this is to take
$g(k,t)=\mathrm{log}\rho(k,t)$ [and, hence,
$f(k,t)=\dot{\rho}(k,t)/\rho(k,t)-\Lambda(t)$] with the new
function $\rho(k,t)$ obeying the auxiliary Ermakov-Pinney equation
\begin{equation}
\ddot{\rho}(k,t)+\Omega^2(k,t)\rho(k,t)=\rho^{-3}(k,t).\label{Ermakov}
\end{equation}
Here the dot denotes $\partial_t$ and
$\Omega^2(k,t)=k^2\omega^2(t)-\Lambda^2(t)-\dot{\Lambda}(t)$, that
we will suppose to be positive in the following. As we can see
this equation is closely related to the classical equation of
motion for a harmonic oscillator with time-dependent frequency
$\Omega(k,t)$. In terms of this new function the state vector
$|\chi(t)\rangle$ satisfies the differential equation
\begin{equation}
i\hbar\partial_{t}|\chi(t)\rangle=\frac{1}{2}\left(\int_{\tilde{\mathbb{X}}}
\frac{1}{\rho^2(k,t)}\left[\hat{\pi}^2(k)+
\hat{\varphi}^2(k)\right]\,\mathrm{d}\mu(k)\right)|\chi(t)\rangle\nonumber
\end{equation}
that can be easily solved to get
\begin{eqnarray}
|\chi(t)\rangle:=
\hat{R}(t,t_{0})|\chi(t_{0})\rangle=\mathrm{exp}\left(-\frac{i}{2\hbar}
\int_{\tilde{\mathbb{X}}}
\left(\int_{t_{0}}^{t}\frac{\mathrm{d}\tau}{\rho^2(k,\tau)}\right)
 \left[\hat{\pi}^2(k)+\hat{\varphi}^2(k)\right]
\,\mathrm{d}\mu(k)\right)|\chi(t_{0})\rangle.\nonumber
\end{eqnarray}
At this point one can build the solution to the original
Schr\"{o}dinger equation from $|\chi(t)\rangle$ and the operator
$\hat{T}(t)$.

The solutions to (\ref{Ermakov}) have some interesting features,
in particular they never vanish and are such that
$|\chi(t)\rangle$ is independent of the initial conditions chosen
for $\rho(k,t)$, as will be shown in appendix \ref{appendixB}.
Other choices for the function $g(k,t)$ allow the formal solution
of the equation for $|\chi(t)\rangle$ but are problematic because
the evolution operators thus obtained are not well defined for all
values of $t$ \cite{FGuasti:2003}. Finally, going back to the
original state vector $|\psi(t)\rangle$, we obtain the formal
quantum evolution operator\footnote{The evolution of a state
vector from time $t_0$ to $t$ is given by
$|\psi(t)\rangle=\hat{U}(t,t_{0})|\psi(t_{0})\rangle$.}
$$
\hat{U}(t,t_{0})=\hat{T}^{\dag}(t)\hat{R}(t,t_{0})\hat{T}(t_{0}),
$$
with $\hat{T}(t)=\hat{S}(t)\hat{D}(t)$ and
\begin{eqnarray}
&&\hspace{-4mm}\hat{D}(t)=\mathrm{exp}\left(-\frac{i}{2\hbar}\int_{\tilde{\mathbb{X}}}
\left[\frac{\dot{\rho}(k,t)}{\rho(k,t)}-\Lambda(t)\right]\hat{\varphi}^2(k)\,\mathrm{d}\mu(k)\right),\nonumber
\\
&&\hspace{-4mm}\hat{S}(t)=\mathrm{exp}\left(\frac{i}{2\hbar}\int_{\tilde{\mathbb{X}}}
\,\,\mathrm{\log}\rho(k,t)\left[\hat{\varphi}(k)\hat{\pi}(k)+\hat{\pi}(k)\hat{\varphi}(k)\right]\,\mathrm{d}\mu(k)\right),\nonumber
\\
&&\hspace{-4mm}\hat{R}(t,t_{0})=
\mathrm{exp}\left(-\frac{i}{2\hbar}\int_{\tilde{\mathbb{X}}}
\left(\int_{t_{0}}^{t}\frac{\mathrm{d}\tau}{\rho^2(k,\tau)}\right)
\left[\hat{\pi}^2(k)+\hat{\varphi}^2(k)\right]
\,\mathrm{d}\mu(k)\right).\nonumber
\end{eqnarray}
It is clear, by construction, that if we can find a suitable
representation for $\hat{\varphi}(k)$ and $\hat{\pi}(k)$ as
self-adjoint operators on a certain Hilbert space,
$\hat{U}(t,t_{0})$ satisfies
\begin{eqnarray*}
&&\hat{U}(t_0,t_0)=\mathbb{I},\\
&&\hat{U}(t,t_0)^{-1}=\hat{U}(t,t_0)^\dag,\\
&&\hat{U}(t_2,t_0)=\hat{U}(t_2,t_1)\hat{U}(t_1,t_0),\\
&&i\hbar\partial_t\hat{U}(t,t_0)=\hat{H}(t)\hat{U}(t,t_0)
\end{eqnarray*}
where $\hat{H}(t)$ is the quantum Hamiltonian (\ref{qHam}). It is
interesting to notice (see appendix \ref{appendixB}) that
$$
\sin\left(\int_{t_{0}}^{t}\frac{\mathrm{d}\tau}{\rho^2(k,\tau)}\right)
=\frac{u(k,t)}{\rho(k,t)\rho(k,t_0)}
$$
where $u(k,t)$ is the unique solution to the homogeneous equation
$ \ddot{u}(k,t)+\Omega^2(k,t)u(k,t)=0$ satisfying $u(k,t_0)=0$ and
$\dot{u}(k,t_0)=1$. This allows us to write explicit and closed
expressions for $\hat{U}(t,t_0)$ whenever we can give closed
solutions to the previous equation for $u(k,t)$ because solutions
to the Ermakov equation can be obtained from these.

In order to discuss the conditions that the representations for
the field and momentum must satisfy it is convenient to factorize
it as
$\hat{U}(t,t_{0})=\hat{\mathcal{D}}(t,t_{0})\hat{\mathcal{S}}(t,t_{0})\hat{\mathcal{R}}(t,t_{0})$
with
\begin{eqnarray*}
&&\hat{\mathcal{D}}(t,t_{0}):=\hat{D}^{\dag}(t)\hat{D}(t_{0}),\\
&&\hat{\mathcal{S}}(t,t_{0}):=\hat{D}^{\dag}(t_{0})
\hat{S}^{\dag}(t)\hat{S}(t_{0})\hat{D}(t_{0}),\\
&&\hat{\mathcal{R}}(t,t_{0}):=\hat{D}^{\dag}(t_{0})\hat{S}^{\dag}(t_{0})
\hat{R}(t,t_{0})\hat{S}(t_{0})\hat{D}(t_{0}).
\end{eqnarray*}
These expressions directly lead to (\ref{D},\ref{S},\ref{R}). The
evolution of the field is
\begin{eqnarray}\hat{\varphi}_{H}(k,t_{0},t)&=&
\rho(k,t)\bigg[\frac{1}{\rho(k,t_{0})}\cos\left(\int_{t_{0}}^{t}\frac{\mathrm{d}\tau}{\rho^2(k,\tau)}\right)
-\rho(k,t_{0})\bigg(\frac{\dot{\rho}(k,t_{0})}{\rho(k,t_{0})}
-\Lambda(t_{0})\bigg)\sin\left(\int_{t_{0}}^{t}\frac{\mathrm{d}\tau}{\rho^2(k,\tau)}\right)\bigg]\,\hat{\varphi}(k)\nonumber\\
&+&\rho(k,t)\rho(k,t_{0})
\sin\left(\int_{t_{0}}^{t}\frac{\mathrm{d}\tau}{\rho^2(k,\tau)}\right)\,\hat{\pi}(k).\label{evolutionphi}
\end{eqnarray}
Similarly we can compute the momentum operator in the Heisenberg
picture
$$
\frac{\mathrm{d}\hat{\varphi}_{H}}{\mathrm{d}t}(k,t_0,t)=-\frac{i}{\hbar}[\hat{\varphi}_H(k,t_0,t),\hat{H}_H(t)]=
\hat{\pi}_{H}(k,t_0,t).
$$
It is straightforward to show that these are formal solutions to
the classic equations of motion --as one would expect given the
quadratic nature of the Hamiltonian-- because the coefficients of
$\hat{\varphi}(k)$ and $\hat{\pi}(k)$ are solutions to the
classical equations
$\ddot{\varphi}(k,t)+\Omega^2(k,t)\varphi(k,t)=0$.

\section{Some properties of the Ermakov-Pinney
equation\label{appendixB}}

We discuss in this appendix some properties of the Ermakov-Pinney
equation
$$
\ddot{\rho}(t)+\Omega^2(t)\rho(t)=\rho^{-3}(t)
$$
that are relevant to show the independence of the evolution
operators from the choice of initial conditions. The most general
analytic solution to this equation\footnote{Here we drop the
parameter $k$ that appears in (\ref{Ermakov}) because the results
are trivially generalized to that case.} can be written as
\begin{equation}
\rho(t)=+\left[Au^2(t)+2Bu(t)v(t)+Cv^2(t)\right]^{1/2}\nonumber
\end{equation}
where $u(t)$ and $v(t)$ are linearly independent solutions to the
time-dependent harmonic oscillator equation
$\ddot{\varphi}(t)+\Omega^2(t)\varphi(t)=0$, $A,B,C$ are constants
satisfying  $AC-B^2=1/W^2[u,v]$, and $W[u,v]:= u\dot{v}-\dot{u}v$
denotes the (time independent) Wronskian of the pair of solutions
$u$ and $v$. The square of a solution can be expressed as
\begin{equation}
\rho^2(t)=[u(t)\,\, v(t)]\left[ \begin{array}{cc} A & B\\ B & C
\end{array}\right]\left[\begin{array}{c} u(t) \\
v(t)\end{array}\right]\nonumber
\end{equation}
where $(u,v)\neq(0,0)$ for all $t$ (zeros of solutions $u(t)$ and
$v(t)$ alternate and never coincide as a consequence of Sturm's
theorem). Hence
\begin{equation}
\left|\begin{array}{cc} A & B \\ B & C
\end{array}\right|=AC-B^{2}=W^{-2}>0,\nonumber
\end{equation}
therefore, the quadratic form $\rho^2(t)$ is positive definite and
$\rho(t) >0$ for all $t$.

The general solution of the time-dependent harmonic oscillator
equation $\ddot{\varphi}(t)+\Omega^2(t)\varphi(t)=0$ can be
expressed as
\begin{eqnarray}
\varphi(t)=\left[\frac{\rho(t)}{\rho_{0}}\cos\left(\int_{t_{0}}^{t}
\frac{\mathrm{d}\tau}{\rho^2(\tau)}\right)-\rho(t)\dot{\rho}_{0}\sin\left(
\int_{t_{0}}^{t}\frac{\mathrm{d}\tau}{\rho^2(\tau)}\right)\right]\,\varphi_{0}
+\rho(t)\rho_{0}\sin\left(\int_{t_{0}}^{t}\frac{\mathrm{d}\tau}{\rho^2(\tau)}\right)\,\dot{\varphi}_{0}
\label{sol}
\end{eqnarray}
where $\rho(t)$ is an arbitrary solution to the associated
Ermakov-Pinney equation
$\ddot{\rho}(t)+\Omega^2(t)\rho(t)=\rho^{-3}(t)$; the subindex $0$
denotes evaluation at $t_0$. Notice that the positivity of
$\rho(t)$ guarantees that the previous expression is well-defined
for any value of $t$. As $\varphi(t)$ is uniquely determined by
the initial conditions $\varphi_{0}$ and $\dot{\varphi}_{0}$ the
functions
$$
\frac{\rho(t)}{\rho_{0}}\cos\left(\int_{t_{0}}^{t}
\frac{\mathrm{d}\tau}{\rho^2(\tau)}\right)-\rho(t)\dot{\rho}_{0}
\sin\left(\int_{t_{0}}^{t}\frac{\mathrm{d}\tau}{\rho^2(\tau)}\right)\quad
\mathrm{and}\quad
\rho(t)\rho_{0}\sin\left(\int_{t_{0}}^{t}\frac{\mathrm{d}\tau}{\rho^2(\tau)}\right)
$$
must be independent of the choice of the solution $\rho(t)$ to the
Ermakov equation. This result has some interesting consequences
regarding the quantum evolution operators studied in the paper.
First of all it is possible to show that the unitary evolution
operator for a single quantum harmonic oscillator with
time-dependent frequency (obtained with the techniques explained
in the paper) is independent of the choice of $\rho(t)$. This is
obvious if one looks at the evolution for the field operator given
by (\ref{evolutionphi}). Second, it is easy to see from
(\ref{sol}) that the solution $u(t)$ to the equation
$\ddot{u}(t)+\Omega^2(t)u(t)=0$ satisfying $u_0=0$ and
$\dot{u}_0=1$ is
\begin{equation}
u(t)=\rho(t)\rho_0\sin\left(\int_{t_0}^t
\frac{\mathrm{d}\tau}{\rho^2(\tau)}\right)\nonumber
\end{equation}
and, hence,
\begin{equation}
\sin\left(\int_{t_0}^t
\frac{\mathrm{d}\tau}{\rho^2(\tau)}\right)=\frac{u(t)}{\rho(t)\rho_0}.\label{fin}
\end{equation}
As the integrand in the previous expression is positive, the
integral is a strictly increasing function of $t$. The value of the
integral itself is given in this case by the unique continuous and
increasing extension to $\mathbb{R}$ of the $\arcsin$ function with
argument $u(t)/[\rho(t)\rho_0]$. This is used to obtain one of the
asymptotic expansions appearing in \ref{GS}.

\section{On the definition of the field $\xi$\label{appendixC}}
We will discuss in this appendix the obtention of the field $\xi$
introduced in section \ref{GS}. The reduced phase space  for the
Gowdy and Schmidt models can be coordinatized
 by $(\phi(k),p_\phi(k))$ and the Hamilton equations derived from (\ref{Ham_Sch_phi_k})
 lead to the time-dependent harmonic oscillator equation
\begin{eqnarray}
\partial^2_T\phi(k,T)+k^2e^{2T}\phi(k,T)=0\,.\label{1}
\end{eqnarray}
It is always possible to write (\ref{1}) in terms of a new field
$\xi$ defined by
\begin{eqnarray*}
\phi(k,T)=h(T)\xi(k,T)\,,\quad  \mathrm{with} \quad h>0\,,
\end{eqnarray*}
to give
\begin{eqnarray}
\partial^2_T\xi(k,T)+2\frac{h'(T)}{h(T)}\partial_T
\xi(k,T)+\bigg(\frac{h''(T)}{h(T)}+k^2e^{2T}\bigg)\xi(k,T)=0\,;
\label{2}
\end{eqnarray}
here $h'$ and $h''$ denote the first and second derivative of the
function $h$ respectively. After performing a time redefinition
$t(T)$, with $t'(T)>0$, the equation (\ref{2}) can be written as
\begin{eqnarray}
\partial_t^2\xi(k,t)+t'^{-2}(T)\bigg(t''(T)+2\frac{t'(T)h'(T)}{h(T)}\bigg)\partial_t\xi(k,t)
+t'^{-2}(T)\bigg(\frac{h''(T)}{h(T)}+k^2e^{2T}\bigg)\xi(k,t)=0\,.\label{3}
\end{eqnarray}
Although (\ref{3}) is not yet a time-dependent harmonic oscillator
equation, it is always possible to eliminate the
($\partial_t\xi$)-term by imposing
\begin{eqnarray}
t''+2\frac{t'h'}{h}=0\,. \label{4}
\end{eqnarray}
The general solution of (\ref{4}) is
\begin{eqnarray}
t(T)=t_0\int_{T_0}^T\frac{\mathrm{d}\tau}{h^2(\tau)}\,, \label{5}
\end{eqnarray}
where $t_0>0$ and $T_0$ are constants of integration. With these
assumptions the equation (\ref{3}) becomes
\begin{eqnarray*}
\partial_t^2\xi(k,t)+\bigg(\frac{h''(T)}{t'^2(T)h(T)}
+k^2\frac{e^{2T}}{t'^2(T)}\bigg)\xi(k,t)=0,
\end{eqnarray*}
with $t(T)$ given by (\ref{5}). This last equation is now of the
form $\partial_t^2\xi(k,t)+\Omega^2(k,t)\xi(k,t)=0$, with
$$
\Omega^2(k,t)=\frac{h''(T)}{t'^2(T)h(T)}+k^2\frac{e^{2T}}{t'^2(T)}\,,
$$
for any choice of $h$.  Finally, it is possible to fix the function
$h$ by requiring that $\Omega^2$ is a sum of $k^2$ plus a function
of time. Explicitly, by demanding $e^{2T}/t'^2(T)=1$, we get
\begin{eqnarray}
t(T)=e^T-e^{T_0}\,,\quad h(T)=\sqrt{t_0}e^{-T/2}\label{6}
\end{eqnarray}
or, equivalently,
\begin{eqnarray*}
T=\log(t+e^{T_0})\,,\quad h=\frac{\sqrt{t_0}}{\sqrt{t+e^{T_0}}}\,.
\end{eqnarray*}
Particularizing (\ref{6}) for $t_0=1$ and $T_0\rightarrow -\infty$
we finally arrive at (\ref{xi})
$$\xi(k,t)=\sqrt{t}\phi(k,t),$$
where the field $\xi$ satisfies
\begin{eqnarray*}
\partial_t^2\xi(k,t)+\bigg(\frac{1}{4t^2}+k^2\bigg)\xi(k,t)=0\,.
\end{eqnarray*}

\end{appendix}


\end{document}